\documentclass[10pt,conference]{IEEEtran}
\IEEEoverridecommandlockouts

\usepackage{amsmath,amssymb,amsfonts}
\usepackage{algorithmic}
\usepackage{graphicx}
\usepackage{textcomp}
\usepackage{cite}
\usepackage{tabularx}
\usepackage{enumitem}
\usepackage{array}
\usepackage{booktabs}
\usepackage[table]{xcolor}  
\usepackage{tcolorbox}      
\usepackage{url}
\usepackage{setspace}
\usepackage{hyperref}       

\begin{document}

\title{From Commits to Confidence: \\Towards Stability-Informed Risk Assessment in Open Source Software
}

\author{\IEEEauthorblockN{ Elijah Kayode Adejumo}
\IEEEauthorblockA{\textit{Computer Science} \\
\textit{George Mason University}\\
Fairfax, USA \\
eadejumo@gmu.edu}
\and
\IEEEauthorblockN{Mariam Guizani}
\IEEEauthorblockA{\textit{Electrical and Computer Engineering} \\
\textit{Queen's University}\\
Ontario, Canada\\
mariam.guizani@queensu.ca}
\and
\IEEEauthorblockN{Brittany Johnson}
\IEEEauthorblockA{\textit{Computer Science} \\
\textit{George Mason University}\\
Fairfax, USA \\
johnsonb@gmu.edu}}

\setstretch{0.9}
\maketitle

\newcommand{\mariam}[1]{\textcolor{blue}{[Mariam: #1]}}
\newcommand{\brittany}[1]{\textcolor{red}{[Brittany: #1]}}
\newcommand{\elijah}[1]{\textcolor{purple}{[Brittany: #1]}}

\begin{abstract}
Open source software (OSS) generates trillions of dollars in economic value and has become essential to the technical infrastructures that power organizations worldwide. As these systems increasingly depend on OSS, understanding the evolution of these projects is critical. While existing metrics provide insights into project health, one dimension remains understudied: project resilience, or the ability to return to normal operations after disturbances such as contributor departures,security vulnerabilities and bug report spikes. We hypothesize that stable commit patterns may serve as an indicator of underlying project characteristics such as mature governance, sustained contributors, and robust development processes, factors that existing research associates with resilience. Our findings reveal that only 2\% of repositories exhibit daily stability, 29\% achieve weekly stability, and 50\% demonstrate monthly stability, while the remaining half are unstable across all levels of granularity. Analysis of the 50 unstable repositories  indicate that 86\% of activity is concentrated among a few maintainers, with the top 3 contributors accounting for over 50\% of commits in the past 5 years. In contrast, the 50 stable repositories distribute work more evenly, with the top 3 contributors representing less than 50\% of commits. Our insights thus far indicate the fragile and multi-dimensional nature of OSS project stability, suggesting a need to go beyond commits to understand how our understanding of stability can be enriched with other considerations such as community engagement metrics and issue or pull request churn. Though our efforts only identified two repositories that achieved stability at all three temporal commit granularities, further investigation into their processes and policies can provide insights and foundations for stability-informed risk assessment in practice.


\end{abstract}


\section{Introduction}
Open source software has been a driving force for increased productivity and innovation across companies and organizations~\cite{nagle2019open, guizani2023rules}. 
Since its early inception, numerous studies have investigated the factors contributing to open source project success~\cite{midha2012factors,margan2015success}, as these projects have become foundational to organizational workflows and products~\cite{guizani2023rules}. 
A recent report by the Open Source Security and Risk Analysis (OSSRA) revealed that approximately 97\% of commercial applications contain open source software~\cite{blackduck2025ossra}. 
This includes safety-critical systems such as medical devices and automotive controllers, which continue to rely on open source software for critical infrastructure, including operating systems and dependency libraries~\cite{sulaman2014development,berntsson2017evaluation}.

The widespread development and use of open source software along with the real potential for negative impact increases the need for a more in-depth   understanding of factors that impact open source product quality.
Prior work has illuminated the centrality of risk management to software product quality~\cite{menezes2019risk} and the impact that people and process quality can have on risk levels and mitigation~\cite{sarigiannidis2014quality}.
However, unique challenges plague open source development that impact the ability to adequately or sustainably manage risk such as contributor churn impacting project momentum~\cite{goggins2021making}\cite{foucault2015impact}, licensing conflicts \cite{cui2023empirical}, and high risk of latent vulnerabilities~\cite{le2024latent}. 

To better understand and support risk management in open source, and drawing from prior work that suggests stable processes can be indicators of product quality~\cite{sarigiannidis2014quality, destefanis2025introducing}, 
we hypothesize that stable open source software development patterns may similarly yield more predictable maintenance and evolution outcomes. 
To evaluate this hypothesis, we leverage the Composite Stability Index (CSI) conceptual framework~\cite{destefanis2025introducing}, which quantifies stability along four dimensions (i.e., commit patterns, issue resolution, pull request processing, and community engagement) to investigate stability in open source development patterns.
In this paper, we present preliminary insights into our investigation, starting with the examination of stable commit patterns across 100 highly ranked repositories.
Our results thus far indicate that only 2\% of repositories remain stable day‑to‑day, with the majority exhibiting stability at a weekly (29\%) or monthly (50\%) level. 
Only two  projects are stable at all three levels of granularity, while half the sample are unstable in all granularity levels. 

Our analysis of commit gains suggests that a metric aiming to predict stability using commits could focus on monthly trends while using weekly trends as early indicators.
We also found that large commit yearly throughput does not necessarily indicate stability in all cases. A classic example from our findings was \textbf{Rust} (29,759 commits/year) shows stability across all granularity, while \textbf{Homebrew-cask} (40,078 commits/year) remains unstable across all granularity.  


Thus, this  paper makes the following contributions:
\begin{enumerate}[label=\roman*.]
    \item \textbf{Multi-granularity stability framework}: We operationalize and extend the Composite Stability Index (CSI) conceptual framework \cite{destefanis2025introducing} by investigating three temporal windows (daily/weekly/monthly) with empirical 
   validation across 100 repositories, revealing that stability is 
   scale-dependent only 2\% achieve daily stability while 50\% 
   achieve monthly stability.
    
    \item 
    \textbf{Empirical discovery of stability-concentration coupling:}
    We establish a statistical association ($\phi$=0.66, OR=24.57) 
    between commit instability and contributor concentration (top-3 
    $>$50\%), suggesting that bus factor and development rhythm are 
    interconnected risk dimensions not captured by existing metrics.

\item 
\textbf{Falsification of volume-stability assumption}: We found that commit throughput alone does not necessarily imply stability (e.g., Homebrew-cask: 40,078 commits/year but unstable in all levels of granularity vs. Rust: 29,759 commits/year and stable in all levels of granularity), challenging conventional activity-based health metrics.

\end{enumerate}

\section{Background \& Related Work}
Many software engineering breakthroughs have been enabled by the availability of open source data from platforms such as GitHub \cite{gousios2012ghtorrent}. 
Commit history, which records the evolution of code and other artifacts in version control systems \cite{blischak2016quick}, has been particularly instrumental in advancing software engineering research. 
By analyzing the nature and patterns of commits, researchers have been able to identify and classify different stages in the software development lifecycle \cite{hattori2008nature}.
Commit history also serves as a valuable tool for team leaders seeking to understand developer productivity \cite{oliveira2020code}. Furthermore, commit-based analyses have been successfully employed to predict fault-prone classes in software systems \cite{chong2018can}, demonstrating the widespread utility of this approach in software engineering literature.

Beyond productivity metrics, commit timestamps have proven useful for understanding developers' work patterns, serving as indicators of potential stress and overtime work \cite{claes2018programmers}. Eyolfson et al. \cite{eyolfson2011time} revealed that commits made between midnight and 4 AM are associated with higher bug rates compared to commits made between 7 AM and noon. Their study also found that developers who commit daily produce less buggy code, suggesting that regular development patterns correlate with code quality.

Commits have also emerged as a key health and sustainability metric in open source software, particularly within the Community Health Analytics Open Source Software (CHAOSS) framework, where they serve as vital indicators of project sustainability \cite{goggins2021open}. More recently, a  research by Destefanis et al. has examined the stability of software repositories \cite{destefanis2025introducing} through the lens of engineering systems and control theory, where bursts of activity followed by dormancy periods signify instability, while more controlled commit patterns indicate stability. 

Building on this foundation, we posit that stable commit patterns could serve as proxy for risk assessment in open-source software. 
To explore this direction, we designed an experiment to answer the following research questions:

\begin{description}

\item[\textbf{RQ$_{1}$}]

\textit{To what extent do highly ranked OSS projects display stable commit rhythms across daily, weekly, and monthly timescales?
}

\item[\textbf{RQ$_{2}$}] \textit{How does stability evolve with coarser aggregations of timescales?}

\item[\textbf{RQ$_{3}$}]
\textit{To what extent do commit-stability scores reveal contributors risks and patterns?}



\end{description}

\section{Methodology}
Our research aims to investigate if and to what extent stable repository patterns can inform risk assessment in open-source software. 
While we acknowledge, and intend to investigate, the variety of potential stability indicators such as stable issue resolution time, pull request merge rates, contributor patterns, and community engagement metrics, in this paper we first focus on commit pattern stability as a foundational component of OSS risk assessment.
To support transparency and replicability, we have made all code, extraction dumps and analysis artifacts available online~\cite{replication_package}.

\subsection{Dataset Selection:}
\label{Data_set_selection}
For our experiments, we selected the top 100 repositories based on the following initial criteria which we found relevant for our research questions and intend to optimize as we build towards the bigger vision:

\begin{enumerate}[label=\roman*.]
\item \textbf{Maturity ($\ge$ 10 years)}: We selected repositories that are at least a decade old, as they provide sufficient historical data to observe stable development patterns and long-term trends.
\item \textbf{Community Interest (Stars $>$ 10,000)}: We prioritized repositories with high star counts as indicators of sustained community interest \cite{borges2018s} and significant user engagement over time.
\item \textbf{Active Adoption (Forks $>$ 9,000)}: We focused on repositories with substantial fork counts, as these represent actual usage and adoption patterns within the developer community \cite{jiang2017and}.
\item \textbf{Software Development Focus}: We excluded educational repositories containing books, programming tutorials, videos, and instructional materials from our analysis to ensure our dataset reflects genuine software development dynamics rather than educational content.
\item \textbf{Active Status}: We excluded archived repositories during manual filtering to maintain a dataset of currently active projects that continue to evolve and receive contributions.
\end{enumerate}


\subsection{Commit Stability Measure (RQ1)}


We empirically evaluate the applicability of the conceptual framework proposed by Destefanis et al. \cite{destefanis2025introducing}. Following inspiration from control theory, they propose a conceptual  framework making the parallel between a stable manufacturing processes yielding more predictable product quality and a stable software development rhythm yielding more predictable maintenance and evolution outcomes. 
They defined the commit frequency function as follows:

\begin{equation}
c(t)=\frac{N_{C}\!\bigl(t,\,t+\Delta t\bigr)}{\Delta t},
\label{eq:commit_freq}
\end{equation}
Where $N_{C}\!\bigl (t,\,t+\Delta t\bigr)$ represents the volume of commit within the time interval $ \!\bigl(t,\,t+\Delta t\bigr)$.  For our experimentation, we defined the interval to be within the last 5 years. A repository is classified as stable \cite{destefanis2025introducing} when:

\begin{equation}
\left|\frac{dc(t)}{dt}\right| \;\le\; \alpha_{c},
\quad \forall\, t \in [t_{0},\, t_{0}+T],
\label{eq:commit_stability}
\end{equation}

The \textbf{parameter $\alpha_{c}$} establishes a maximum threshold for acceptable fluctuations in commit activity over a period $ [t_{0},\, t_{0}+T]$. 
\textbf{When this threshold is surpassed, the repository loses its stable classification}. 
The boundary value $\alpha_{c}$ \cite{destefanis2025introducing} is specified as:

\begin{equation}
\alpha_{c} = \frac{\sigma_{\text{daily commits}}}{\mu_{\text{daily commits}}} \le 0.5.
\label{eq:threhold limit}
\end{equation}

\paragraph{Granularity variants}
To examine stability at different levels of granularity, we
instantiate Eq.~\eqref{eq:commit_freq} from prior work~\cite{destefanis2025introducing} at three window sizes:
\[
\Delta t=
\begin{cases}
1\text{day}   &\!\!: \; c_{d}(t) \\
7\text{days}  &\!\!: \; c_{w}(t) \\
30\text{days} &\!\!: \; c_{m}(t)
\end{cases}
\]
and re‑apply the stability test in
Eq.~\eqref{eq:commit_stability} with the same coefficient‑of‑variation
threshold (Eq.~\eqref{eq:threhold limit}, $\alpha_{c}\!=\!0.5$) at each granularity.
\textbf{A repository is therefore labeled
\emph{daily‑stable}, \emph{weekly‑stable}, or \emph{monthly‑stable}
when the inequality holds for $c_{d}(t)$, $c_{w}(t)$, or $c_{m}(t)$,
respectively}.
The three labels allow us to study how stability classifications
vary across granularity.

\par\vspace{0.7\baselineskip}
\paragraph {Target normalizer for the commit metric}
\label{Target Normalizer function}
Destefanis et al. ~\cite{destefanis2025introducing} map each raw stability signal onto $[0,1]$ with the \emph{triangular normaliser} in the CSI framework this mapping is applied to four components
$\{c,i,p,a\}$—\emph{commit frequency, issue resolution,
pull‑request merge rate,} and \emph{community activity}.

In this study, we focus on the \textbf{commit component}
($k=c$).  
Following the original thresholds, we set the
\emph{target} and \emph{tolerance} to
$\mu_c = 0.25$ and $\sigma_c = 0.25$; that is, we consider a
Coefficient‑of‑Variation (CV) of~0.25 to be ideal, and we tolerate
deviations up to $\pm0.25$:

\[
\phi_c(x)=
\begin{cases}
1 - \dfrac{|x-0.25|}{0.25}, & 0 \le x \le 0.50,\\[6pt]
0, & \text{otherwise.}
\end{cases}
\]

Thus a repository with monthly $\text{CV}=0.25$ receives the maximum stability score $\phi_c=1$, while any CV outside the corridor
$[0.00,\,0.50]$ is mapped to $\phi_c=0$ (treated as unstable at that
granularity).

\subsection{Stability evolution across granularities (RQ2)}
To assess how the stability score evolves after applying the target normalizer described in Section~\ref{Target Normalizer function} across the three granularity (daily, weekly, monthly), we evaluate the rate of change function. That is for every repository i we hold three stability scores:
$\phi_{d,i}, \phi_{w,i}, \phi_{m,i}$, where $\phi_{d,i} =\text{daily stability score}$, $\phi_{w,i} = \text{weekly score}$, and  $\phi_{m,i}= \text{monthly score}$  


To reveal the stability evolution as the window widens, we compute two stepwise deltas:

Step 1: Daily $\rightarrow$ Weekly, 
\begin{align}
\Delta_{dw,i} &= \phi_{w,i} - \phi_{d,i}
\end{align} This reveals how much the stability score rises (or falls) when we aggregate commits from 1 day into 7‑days. 

Step 2: Weekly $\rightarrow$ Monthly
\begin{align}
\Delta_{wm,i} &= \phi_{m,i} - \phi_{w,i}
\end{align}
This reveals additional change when we further aggregate those 7‑day window  into a 30‑day window.

\subsection{Commit Stability and Contributor Patterns (RQ3)}

To examine how contributor patterns impact commit-stability measures in practice and vice versa, we computed the percentage of commits made by each repository's top three contributors over the last five years. 
The ``bus factor''  measure \cite{cosentino2015assessing} approximates team resilience, that is, \textbf{a high share of commits concentrated in the top three contributors indicates a low bus factor and a potential risk factor}, where the departure of a few maintainers could critically impact repository continuity.
This process also involved excluding commits made by bots or automated CI/CD services such as release and dependency-update bots. We inspected contributor identities to ensure that contributor names corresponded to human accounts rather than bots, allowing us to filter out automated activity from our analysis.

We tested the association between monthly stability (stable vs. unstable) and maintainer centralization (top-3 $\ge$50\% vs. $<$50\% of commits) using a Pearson chi-square test of independence on a 2×2 contingency table. We report $\chi$², p-values, and $\phi$  as an effect size, as well as odds ratios with 95\% confidence intervals.





\section{Preliminary Findings}
Our efforts thus far have provided valuable initial insights into stability-driven risk assessment. 
Below we detail our findings regarding daily, weekly, and monthly commit patterns, stability across these levels of granularity, and relationship to contributor patterns.

\subsection{Stable Commit Rhythms (RQ1) }

Our analysis revealed a hierarchy in commit rhythm stability where only 2\% of repositories (specifically \textsc{Rust} and  \textsc{nixpkgs}) maintain consistent daily patterns, 29\% achieve stable weekly rhythms, and half (50\%) demonstrate stable monthly commit patterns. 
This suggests that most projects find their natural cadence at coarser timescales. 
All 27 repositories that are stable at the weekly scale remain stable at the monthly scale, whereas 21 repositories achieve stability only at the monthly granularity. 
Table~\ref{tab:Repository stability profiles across the three granularity} highlights the cross‑granularity overlap.

\begin{table}[ht]
\centering
\caption{Repository stability profiles across the three levels of granularity} 
\label{tab:Repository stability profiles across the three granularity}
\begin{tabular}{lc}
\toprule
\textbf{Stability profile (Granularity)}      & \textbf{Repos} \\
\midrule
Stable at all \textbf{three} scales (daily–weekly–monthly)   & 2  \\
Stable at \textbf{two} scales (weekly–monthly)               & 27 \\
Stable at \textbf{one} scale (monthly only)                  & 21 \\
Unstable at \textbf{all} three scales                        & 50 \\
\bottomrule
\end{tabular}
\end{table}

Notably, half of all repositories show no stable rhythm at any timescale, indicating highly irregular development patterns. 
In contrast, only 2 repositories  maintain the discipline of consistent daily commits while preserving stability at weekly and monthly scales representing true stability at all levels of granularity. 
This hierarchy suggests that while many projects can establish longer-term release cadences, maintaining day-to-day consistency could require exceptional efforts. The finding that 29\% of repositories achieve weekly stability aligns with common sprint-based development practices, while the 50\% achieving monthly stability may reflect broader release cycles or less structured development workflows.

\subsection{Stability Evolution Across Timescales (RQ2) }

Our evaluation of stability score evolution across the 50 identified stable repositories revealed that the two repositories with stability across all timescales (\textsc{Rust} and \textsc{nixpkgs}) differ in evolution with changes in granularity.
\textsc{Rust}'s stablity score declines at coarser granularity (daily to monthly), while \textsc{nixpkgs}' score improves at coarser granularity. 
This implies that coarser aggregation may not always smooth out irregularities.
The stability scores are shown in Table \ref{tab:stability_profiles}.

\begin{table}[ht]
\centering
\caption{Evolution in stability across granularity} 
\label{tab:stability_profiles}
\begin{tabular}{lccc}
\toprule
\textbf{Repository}  & \textbf{Daily} $\phi_c$ & \textbf{Weekly} $\phi_c$  & \textbf{Monthly} $\phi_c$ \\
\midrule
rust‑lang/rust & 0.91 & 0.68 & 0.52 \\
NixOS/nixpkgs & 0.27 & 0.51 & 0.62 \\

\bottomrule
\end{tabular}
\end{table}

We also observed the evolution among the 29 repositories stable at the weekly granularity by tracking how their scores changed at the monthly scale. 
For twenty‑six of them ($\approx 90\%$) their stability score increased, while three experienced a modest decline. 
Thus, broadening the window from weekly to monthly can smooth  residual burstiness, but not always. 
Detailed descriptive statistics for the weekly‑to‑monthly deltas are summarized in Table \ref{tab:Weekly-Monthly Stability Evolution}.
Monitoring both granularity would be beneficial for nuanced insight.

\begin{table}[ht]
\centering
\caption{Weekly-Monthly Stability Evolution} 
\label{tab:Weekly-Monthly Stability Evolution}
\begin{tabular}{lc}
\toprule
\textbf{Metric}      & \textbf{Value} \\
\midrule
 Stability improves   & $26/29$ ($\approx 90\%$) \\
Stability degrades  & $3/29$ ($\approx 10\%$)\\
Mean/ Median change  & + 0.33 / + 0.34 \\
Largest improvement  & + 0.79 \\
Largest degrade & - 0.16 \\
\bottomrule
\end{tabular}
\end{table}

\subsection{Commit Stability and Contributor Patterns (RQ3)}
\label{Commit_Stability_and_Contributor}
  
We tested whether commit stability can reflect or predict team resilience by examining contributor concentration. 
We observed that 86\% (43/50) of unstable repositories exhibited high concentration of top-three contributor contributions, with the top three contributors responsible for $\ge$50\% of all commits in the last five years.
Stable repositories exhibited more distributed contributions where  we found that for 80\% (40/50) of repositories the top three contributors were responsible for $<$50\% of commits over the same period.
To better understand these distributions, we examined the actual share of commits made by the top three contributors. 
Among the 40 stable repositories, the median top-3 share was 34.29\% (mean 31.64\%). In contrast, among the 43 unstable repositories, the median top-3 share was 80.70\% (mean 78.65\%).  
A detailed statistical analysis is summarized in Table \ref{tab:centralization_embedded_stats}

\textsc{Rust} and \textsc{nixpkgs}, which were stable across all three levels of granularity, further illustrate this pattern. 
In \textsc{Rust}, the top three contributors account for only 18.44\% of all commits in the last five years, and in \textsc{nixpkgs}, the top three contributors account for 28.99\%. 
This suggests that highly stable projects tend to combine regular commit rhythms with broadly distributed contributions.

\begin{table}[t]
\centering
\caption{Top contribution patterns vs our monthly stability measure}
\label{tab:centralization_embedded_stats}
\begin{tabular}{lrrr}
\toprule
& \textbf{Unstable} & \textbf{Stable} & \textbf{Total} \\
\midrule
Top-3 $\geq$ 50\% & 43 & 10 & 53 \\
Top-3 $<$ 50\%    &  7 & 40 & 47 \\
\midrule
\textbf{Total}    & 50 & 50 & 100 \\
\midrule
\multicolumn{4}{l}{\textbf{Test statistics}} \\
$\chi^2$ (df=1)     & \multicolumn{3}{r}{43.72} \\
$p$-value           & \multicolumn{3}{r}{$<0.0001$} \\
$\phi$ (Cram\'er’s $V$) & \multicolumn{3}{r}{0.66 } \\
Odds ratio [95\% CI] & \multicolumn{3}{r}{24.57 \ [8.53, 70.74]} \\

\bottomrule
\end{tabular}

\end{table}

\section{Discussion \& Future Work}

Our findings introduce a new dimension for assessing open source software risk, extending beyond technical considerations to the consideration of social factors. 
Though our insights are early, they illuminate signals that suggest  stable commit patterns may indicate underlying project characteristics such as mature governance, sustained contributors, and robust development processes, factors that existing research associates with resilience when facing disruptions \cite{malgonde2023resilience}.

\subsection{The Conundrum of OSS Stability}

Maintainers and contributors are central to open source software success \cite{dias2021makes,milewicz2019characterizing}, yet they face significant challenges including stress and burnout from keeping up with project demands~\cite{geiger2021labor,raman2020stress}.
Our findings provide additional clarity into the role of contributors in open source project  stability. 
The majority of unstable projects show a low bus factor where the project is dependent on a small group of  top contributors. This concentration of contribution creates a high risk context where the unavailability of one contributor is likely to generate disruption. 
At the same time, the majority of stable repositories are maintained by a more balanced set of contributors. 
This suggests that coarse grained stability assessments can reflect genuine resilience rather than masking project fragility.
The strong association found in our efforts between contributor concentration and instability reveals maintainer stress patterns identified in prior work \cite{geiger2021labor,milewicz2019characterizing}.This aligns with longstanding efforts to attract and retain OSS contributors, which have proven challenging. Difficulties such as onboarding and supporting newcomers are unique to the open source context and, if unaddressed, can become inherent barriers to project stability \cite{bouktif2014predicting}.

Our insights thus far suggest an important direction for future work is acquiring a deeper understanding of the unique barriers to stability in open source and mechanisms that can support practical and realistic stability measurement and prediction.
Our proposed stability-informed assessment offers an actionable novel approach for projects to monitor their commit and contributor patterns to identify unhealthy trends before they escalate.
For example, projects showing both high concentration and declining stability scores warrant immediate intervention, as they exhibit compounded risk of maintainer burnout and project abandonment.
This metric could trigger proactive community interventions, such as recruiting additional contributors or encouraging maintainers to take breaks that would not have significant impact on project stability. 
By providing quantitative indicators of development stress patterns, this approach could enable data-driven community health management.





\subsection{Characterizing the Relationship between Stability \& Risk}

While in some software contexts software availability and fit may be sufficient for adoption or use, in many others (e.g., safety critical software~\cite{sulaman2014development,berntsson2017evaluation}) a key risk involved in risk assessment is whether the software will be reliably maintained over time or in the long term.
Our findings suggest that commit patterns, and the contributor patterns underlying them, can provide meaningful insights into the fragility of a given project.

While commit patterns identified in our efforts thus far provide valuable insights, they represent only one dimension of project health. 
A comprehensive stability-informed risk assessment requires multiple indicators, as suggested by \cite{destefanis2025introducing}, including issue resolution times, pull request merge rates, and community engagement metrics. 
Future work will extend our stability analysis framework to incorporate these additional signals, including GitHub Discussions board utilization and engagement patterns \cite{hata2022github}.


It is also important to acknowledge that risk is multi-faceted and goes beyond contributor risk. 
Central to risk is the introduction of vulnerabilities, a risk that increases with contributor churn 
\cite{bosu2014identifying}.
One promising direction for future work is to examine how commit stability and contributor centralization relate to security vulnerability profiles. Our preliminary findings distinguish stable, distributed projects from unstable, highly centralized among few contributors, the next step is to investigate whether these structural patterns translate into different security behaviors.  
An example could be evaluating whether unstable, centralized repositories accumulate more vulnerabilities, experience higher-severity Common Vulnerabilities and Exposures (CVEs), or exhibit longer time-to-fix windows than stable distributed repositories especially around periods where stability scores drop. If such links hold, commit stability and contributor concentration could serve as lightweight early warning indicators for dependency risk and incident response capacity in safety-critical settings.




\subsection{Limitations}
Our current analysis identifies correlations and associations but does not yet show that stability predicts response capacity during real disruptions. 
Likewise, the association between stability and contributor distribution does not imply causation; both may be driven by organizational support, funding levels, or domain criticality. Because our sample focuses on highly ranked, mature repositories (section \ref{Data_set_selection}), the findings may not generalize to smaller, younger, or less popular projects. These motivate future work to test stability as a predictive signal and to extend the analysis across more diverse project scales.


\subsection{Future Work}

Contributors are essential to OSS success and sustainability, yet identifying suitable contribution opportunities remains challenging \cite{steinmacher2015social,alderliesten2021initial}. Our proposed stability-informed assessment offers a novel approach to this problem by highlighting projects with stable commit patterns, which may indicate mature development processes and greater receptiveness to new contributors.
Projects demonstrating consistent, predictable development rhythms often have established workflows, clear governance structures, and active maintainer engagement, factors that facilitate successful contributor onboarding \cite{adejumo2024towards}.

Our analysis identified 50 monthly-stable repositories (50\% of our dataset), including 29 that achieve weekly stability and 2 that maintain 
daily consistency. Our next phase involves systematic comparative analysis across this stable cohort to identify common organizational practices and project characteristics that enable consistent development rhythms.
The qualitative analysis will include \textbf{document analysis} (e.g., public governance documents, contribution guidelines, CI/CD workflow definitions, and release policies) and  
\textbf{maintainer interviews} to  better understand how stability-enabling structures. 
As a part of this effort, we will also investigate the relationships between project characteristics and coarse or fine grained stability.
We believe this comparative approach will reveal whether stable repositories share 
transferable organizational patterns and to what extent stability results from 
context-specific factors (e.g., funding, domain criticality, team culture). 

Since our overarching argument is that stable development processes can yield more reliable software, a natural next step is to relate commit stability and contributor centralization to static-analysis-based fault-proneness metrics. In future work, we also plan to compare stable and unstable repositories on measures such as cyclomatic complexity, code smells, warning density, and change-prone ``hotspot'' files, and to examine whether unstable, highly centralized projects exhibit systematically more fault-prone code than stable, more distributed ones.
By systematically expanding our insights on stability in open source, we can continue to make progress towards support for stability-informed risk assessment in practice.

\bibliography{references}

\end{document}